\documentclass{openjournal}

\usepackage{lipsum}

\usepackage{xcolor}
\usepackage{textgreek}
\usepackage[utf8]{inputenc}
\usepackage[english]{babel}

\usepackage{hyperref}
\hypersetup{
    unicode, 
    colorlinks=true,
    linkcolor=linkcolor,
    citecolor=linkcolor,
    filecolor=linkcolor,
    urlcolor=linkcolor,
}
\usepackage{color,colortbl}
\definecolor{linkcolor}{rgb}{0.0,0.3,0.5}
\usepackage{tensind}
\tensordelimiter{?}
\DeclareGraphicsExtensions{.bmp,.png,.jpg,.pdf}
\usepackage{verbatim}
\usepackage[normalem]{ulem}
\usepackage{orcidlink}
\usepackage{soul}

\begin{document}
\title{Galactic Contrail in NGC 3627 caused by Dwarf Galaxy Candidate or Massive Black Hole Flyby}



\author{Mengke Zhao\orcidlink{0000-0003-0596-6608}}
\email{mkzhao@nju.edu.cn}
\affiliation{School of Astronomy and Space Science, Nanjing University, 163 Xianlin Avenue, Nanjing 210023, Jiangsu, People’s Republic of China}
\affiliation{Key Laboratory of Modern Astronomy and Astrophysics (Nanjing University), Ministry of Education, Nanjing 210023, Jiangsu, People’s Republic of China}

\author{Guang-Xing Li\orcidlink{0000-0003-3144-1952}}
\email{gxli@ynu.edu.com}
\affiliation{South-Western Institute for Astronomy Research, Yunnan University, Kunming 650091, People’s Republic of China}

\begin{abstract}
We report the discovery of a kiloparsec-scale molecular contrail in the spiral galaxy NGC 3627—a narrow structure spanning $>8$ kpc in length with a width of $\sim200$ pc and an extreme aspect ratio $>40$, observed in both mid-infrared dust emission (PHANGS-JWST) and CO(2-1) gas (PHANGS-ALMA). 
This contrail size significantly exceeds the size of any known analogues in the Milky Way and exhibits supersonic turbulence ($\sigma_v \approx 10\ \mathrm{km~s^{-1}}$). 
Its morphology and dynamics are consistent with gravitational focusing by a
flyby compact object of mass $\gtrsim 2 \times 10^6~M_{\odot}$, likely a
massive black hole or a dwarf galaxy nucleus, traversing the
disk at $ \gtrsim 300~\mathrm{km~s^{-1}}$. The crossing time of such a contrail,
estimated from its width and velocity dispersion, is only $\sim 20$ Myr,
implying a recent interaction. This contrail can be caused by a dwarf galaxy, or
massive black hole nucleus. This discovery establishes galactic-scale contrails
as probes of massive dark objects interacting with medium in and around galactic disks.
\end{abstract}

\begin{keywords}
    {galaxy, dwarf galaxy, contrail}
\end{keywords}

\maketitle

\section{Introduction}
\label{sec:intro}

Galaxies are basic ingredients of the universe  \citep{2016ARA&A..54..667S,2022ARA&A..60...73S}, consisting of complex systems with stars, gas, and dark matter. 
They are considered the main sites for star formation in the universe. 
In galactic discs, gas is organized into spiral arms with significant curvature,
primarily driven by stellar gravitational potentials arising from disc
instabilities  \citep{2012ApJ...756...45L}, where stellar density waves generate
asymmetric gravitational potentials that torque and compress gas, forcing it
into coherent, curved trajectories
\citep{1969ApJ...158..123R,1973PASA....2..174K,2016ARA&A..54..667S}. 

Molecular contrails represent coherent linear structures in the interstellar medium
(ISM) with high aspect ratios
\citep{2021MNRAS.503.4466L,2022MNRAS.511..980S,2023ApJ...945...39K,2023ApJ...946L..50V,2024MNRAS.527.5503O}.
They are unlikely to form through typical galactic processes nor turbulence, and
a promising formation mechanism is gravitational focusing caused by passages of compact objects (e.g., massive stars, black holes
\citep{2021MNRAS.503.4466L,2023ApJ...945...39K,2023ApJ...946L..50V,2024MNRAS.527.5503O},
or star clusters  \citep{2021MNRAS.503.4466L}, even dwarf galaxies
\citep{2023MNRAS.524.1431Z}) through ambient gas.


The population of compact systems around galaxies is diverse, and our knowledge
about them remains limited.
Many dwarf galaxies exist as satellite systems around host galaxies  \citep{2024ApJ...976..119W}. 
While large dwarf galaxies (e.g., the Large Magellanic Cloud around the Milky Way) are readily observable  \citep{2021ApJ...920L..19S,2024ApJ...967...72R}, ultra-faint dwarfs remain extremely challenging to detect directly and often require indirect methods  \citep{2024ApJ...976..117M,2025ApJ...979..164C}. 
Detection of Milky Way satellites proves particularly difficult
\citep{2012A&A...544A.113W,2024ApJ...976..119W}, though their presence in
external galaxies like NGC 3627 may be inferred through gravitational
interactions. Recent JWST observations have unveiled the little red dots \citep{Geraldi2024,Marconcini2025,Akins2024,Labbe2025}, which
are considered compact, red-tinted, early-universe galaxies, likely powered by
supermassive black holes or dense star-forming regions. The discovery of these
little red dots points to a gap in our understanding of the population of
compact systems around galaxies, and kpc-scale contrails serve as unique indirect
tracers for such systems.

\section{Galactic-scale contrail in extra galaxy NGC 3627}
\subsection{Data}
\begin{figure}
    \centering
    \includegraphics[width=\linewidth]{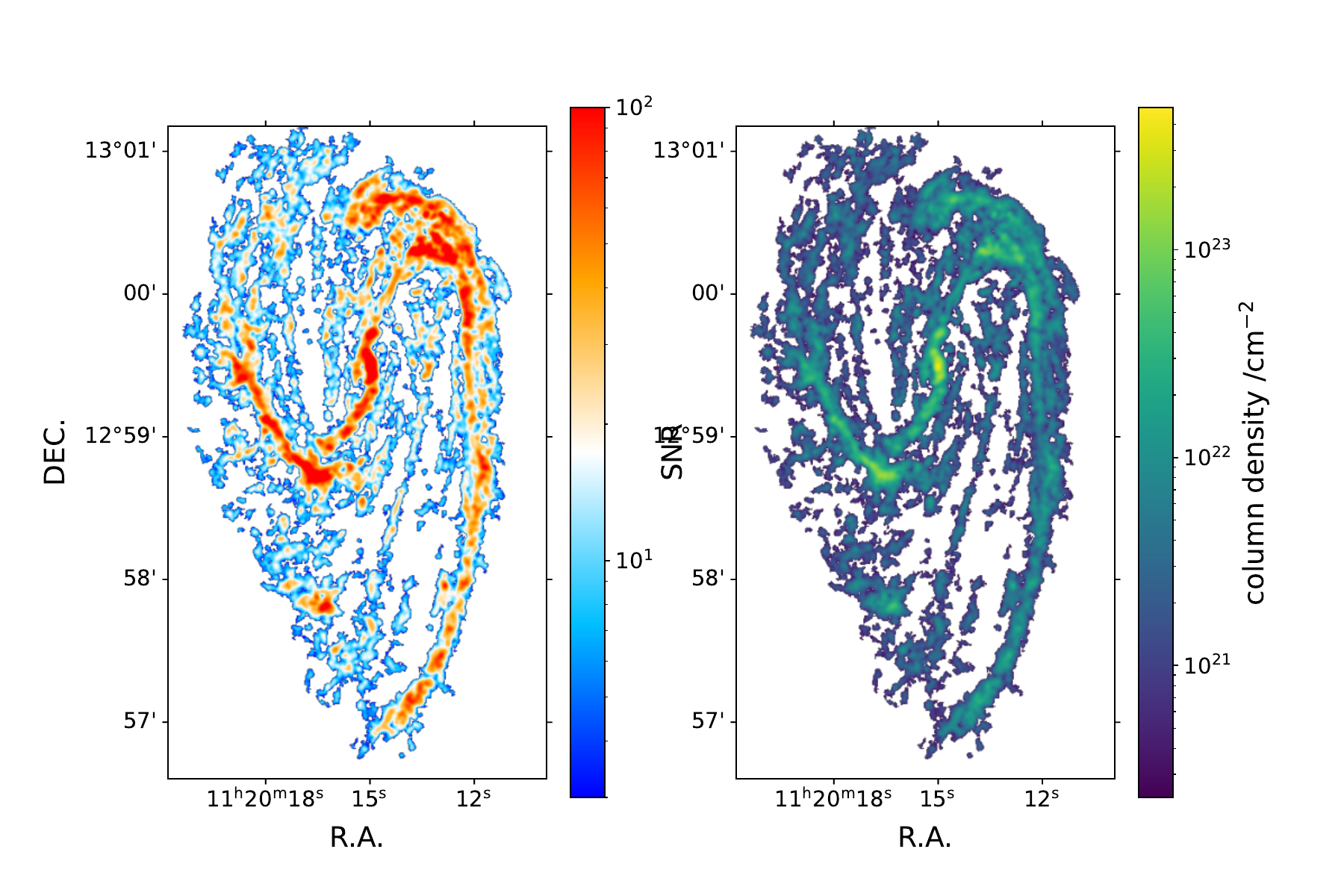}
    \caption{{\bf The signal-to-noise ratio of CO 2-1 and the column density distribution.}
    The left panel shows the signal-to-noise ratio of CO 2-1 spectral line in NGC 3627.
    The right panel displays the distribution of column density derived from CO 2-1 intensity map ($\Sigma$ = $I_{\rm CO 2-1} R_{\rm 2-1}^{-1} X_{\rm CO}$, 
    where the $R_{\rm 2-1}$ is the CO(2-1)/CO(1-0) ratio \citep{2021MNRAS.504.3221D} and the $X_{\rm CO}$ is the CO-to-H$_2$ conversion factor around 2$\times$10$^{20}$ (K km s$^{-1}$)$^{-1}$ \citep{2013ARA&A..51..207B}.}
    \label{figA1}
\end{figure}

NGC 3627 is a barred spiral galaxy in Leo, lies 31 million light-years away and exhibits active star formation. ALMA observations from the PHANGS survey map its molecular gas clouds, revealing an asymmetric structure \citep{leroy2021, sun2023}.
We select $^{12}$CO\,(2-1) emission as the main tracer to measure the gas
content.
The spectral cube data comes from the PHANGS-ALMA survey  \citep{2021ApJS..257...43L}, which includes 90 nearby galaxies (d $\leq$  20\,Mpc). 
We use the signal-to-noise ratio and column density map to present the detail of the PHANGS-ALMA CO spectral in NGC 3627 (see Fig.\,\ref{figA1}).
The dust emission at the wavelength of infrared band come from the PHANGS-JWST survey  \citep{2023ApJ...944L..17L}
The magnetic field orientation is observed by the VLA at 8.46 GHz with a resolution of 11 $''$ using the synchrotron polarization \citep{2001A&A...378...40S}.

\label{sec:sec}

\begin{figure}
    \centering
    \includegraphics[width=\linewidth]{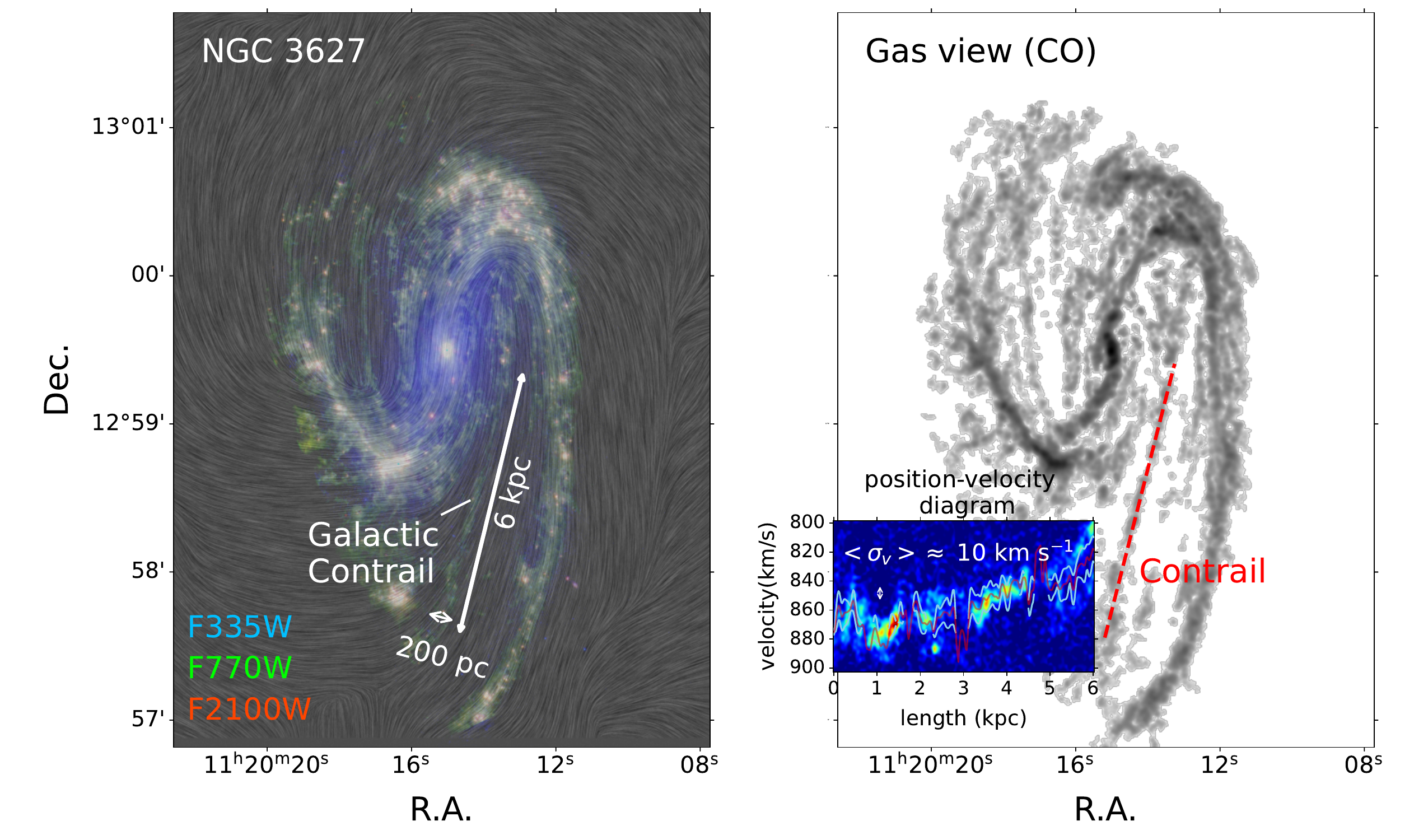}
    \caption{{Galactic-scale Contrail in NGC 3627.}
    The left panels show an RGB image from mid-infrared emission, observed by JWST (red: F2100W, green: F770W, and blue: F335W).
    The write arrows display the cover area of galactic-scale contrail, whose length is around 6\,kpc and width is around 200\,pc.
    The line integral convolution map shows the magnetic field morphology observed with the VLA at the resolution of 11 arcsec \citep{2001A&A...378...40S}.
    The right panel shows the gas view of the NGC 3627 and this contrail, derived by CO(2-1) emission from the PHANGS-ALMA survey \citep{2021ApJS..257...43L}. 
    The sub-panel shows the position-velocity diagram from CO spectral line along the contrail elongation from south to north.
    The red line and blue lines shows the main structure of PV diagram and its velocity dispersion along the contrail long axis, which is fitting a Gaussian model to obtain the mean and standard deviation at each unit length.
    The mean velocity dispersion along the contrail is around 10 km\,s$^{-1}$, whose size is shown as the white arrow in the PV diagram.
    The detail of the data is shown in Appendix. \ref{data}.}
    \label{fig1}
\end{figure}

\begin{figure}
    \centering
    \includegraphics[width=\linewidth]{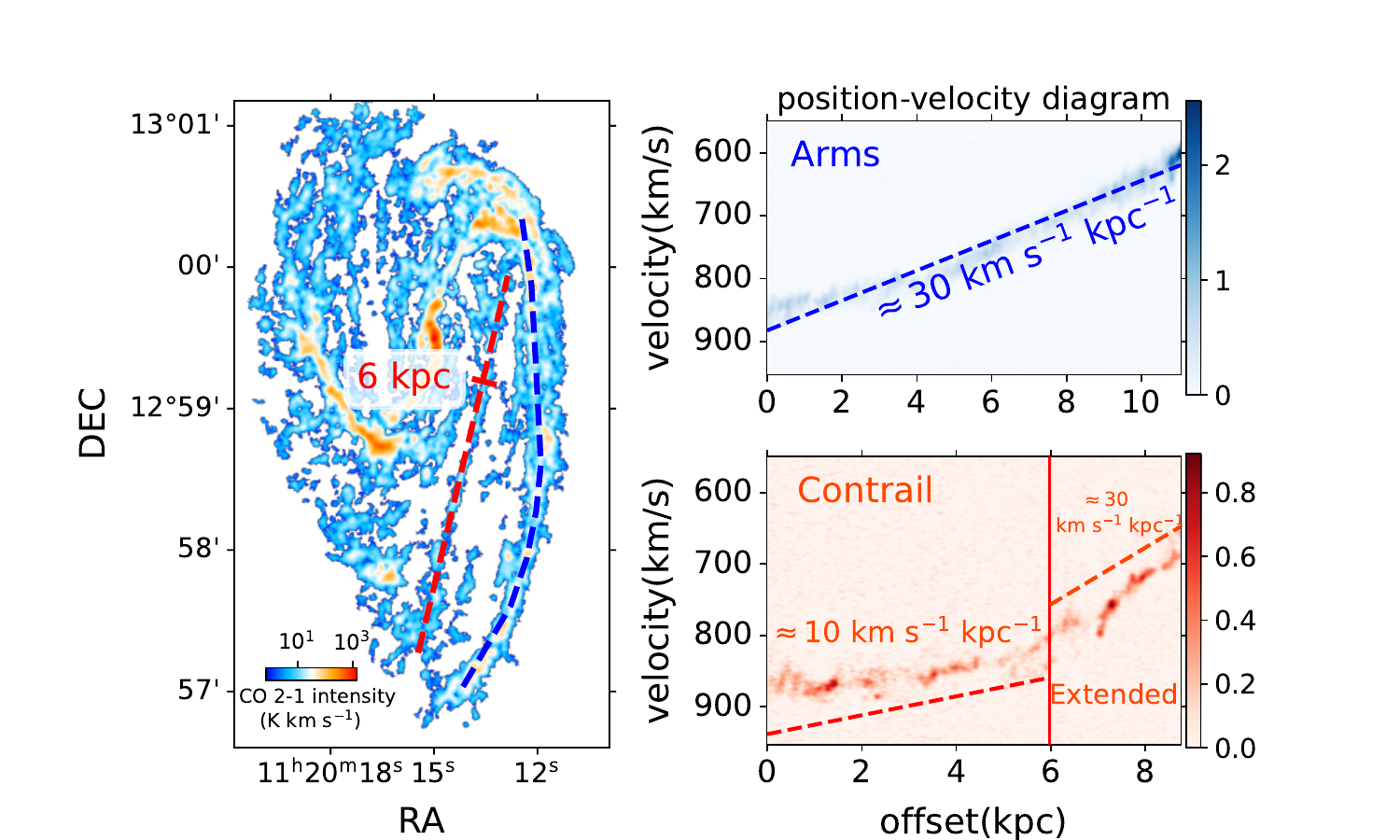}
    \caption{{\bf Velocity structure of galactic-scale contrail, its extended region and spiral arms}.
    The left panel presents the galaxy structure of NGC 3627, probed by the CO 2-1 spectral line.
    The red and blue dot lines show the structural skeleton of galactic-scale contrail, its extended region, and spiral arm.
    The right panels display the position-velocity diagram along the structural skeleton.
    The arms and the extended region of the contrail have a similar velocity gradient around 30\,km\,s$^{-1}$\,kpc$^{-1}$, where the extended region of the contrail exceeds 6 kpc and is close to spiral arms.
    The boundary between the contrail and its extended region is shown as red lines in PV diagram and intensity map on the plane of sky.
    The contrail, inner 6\,kpc, has a flat velocity gradient around 10\,km\,s$^{-1}$\,kpc$^{-1}$.
    }
    \label{fig2}
\end{figure}



\subsection{Galactic-scale contrail in NGC 3627}
An intriguing narrow, straight ISM structure is observed in the external galaxy NGC\,3627.
As shown in Fig.\,\ref{fig1}, this structure contains both gas, traced by CO, and dust, traced by JWST F770W emission. 
In contrast to other regions of the galaxy, it exhibits lower star formation activity, as probed by F2100W, and weaker emission from young stellar objects, as indicated by F335M.
This structure extends at least 6\,kpc in length (a lower limit due to
projection effects) while being only 200\,pc wide (see Fig.\,\ref{fig1}), whose
aspect ratio exceeds 30. It likely represents a galactic-scale contrail
\citep{2021MNRAS.503.4466L,2022MNRAS.511..980S,2023ApJ...945...39K,2023ApJ...946L..50V,2024MNRAS.527.5503O},
whose dimension far exceeds that of any known
small-scale contrails in the Milky Way. 

The velocity structure of this galactic-scale contrail is unique and different from the spiral arms and even its own extended region.
As Fig.\,\ref{fig2} shows, the gas within the main 6-kpc contrail has a very flat velocity gradient along its length, only about 10 km/s per kpc. 
This is much lower than the gradient along the adjacent spiral arm, which is about 30 km/s per kpc.
However, the extended region of the structure, which lies closer to the spiral arm and the galactic bar, has a steeper velocity gradient similar to the arm's. 
The 30 km\,s$^{-1}$ kpc$^{-1}$ gradient matches the galaxy's angular frequency
(Ω = Vc/R) at that distance from the center, as calculated from its circular
velocity \citep{2003A&A...405...89C,2017A&A...597A..85B}. This indicates that this extended region is more affect by the galaxy's shear
compared against the main part of the contrail.



\section{Flyby-induced contrail formation in  NGC3627}

We propose that this striking feature, with a width of 200 pc and a velocity dispersion of $\sim$10\,km\,s$^{-1}$, is the result of a 
flyby event, where an object, possibly an massive black hole or the nucleus of a dwarf galaxy, traversed the galactic disk.

\paragraph {Formation Mechanism: From Warm Gas to Dense Molecular Contrail}

The formation of this galactic-scale contrail is attributed to the momentum injected into the ISM as the massive object crossed the galactic disk through gravitational force \citep{2021MNRAS.503.4466L}. This compressed the warm neutral medium (HI gas, T $\approx$ 8000 K;  \citealt{1969ApJ...155L.149F,2003ApJ...586.1067H,2003ApJ...587..278W}), which typically fills galactic disks  \citep{2008AJ....136.2563W}. 
The compression triggered a rapid cooling process, transforming the warm neutral gas into dense molecular gas, thereby forming the collimated contrail structure  \citep{2024MNRAS.527.5503O}.
Evidence supporting this shock-driven cooling mechanism comes from the observed turbulent velocity of the contrail, measured at $\sim$ 10 km s$^{-1}$ from CO spectral lines (see P-V diagram in Fig.\,\ref{fig1}).
This velocity exceeds the sonic speed of warm HI ($\sim$ 7.5 km s$^{-1}$), confirming the presence of supersonic dynamics consistent with a shock-driven event. 
Since the infall velocity is expected to exceed the sound speed of the ambient
medium, we expect the passage dense object to trigger the formation of the cold
molecular gas, and this is consistent with the detection of the CO molecules in
the contrail, where gas should also reach the self-shielding limit \cite{2014ApJ...790...10S}.
Furthermore, magnetic fields, probed via synchrotron polarization, are found to align with the contrail \cite{2001A&A...378...40S,2023MNRAS.519.1068L}. This contrail-magnetic field alignment can be the result of contraction of the contrail.

\paragraph {Dynamical age and stability}

Since differential rotation (shear) and turbulent motions can disrupt such intricate
structures, we expect them to be produced in a recent event, so short that
neither shear \citep{2008gady.book.....B}  nor turbulence
\citep{2014PASA...31...35D,2021MNRAS.503.4466L} has had time to take effect.

From the width (200 pc) and velocity dispersion ($\sim$10 km\,s$^{-1}$) of the
contrail, we estimate a crossing time of
\begin{equation}
    t_{\rm cross} \approx \frac{d}{\sigma_v} \approx \frac{200\,{\rm pc}}{10\,{\rm km\,s^{-1}}} \approx 20\,{\rm Myr}\;.
\end{equation}

The crossing time is also the time for the contrail to disperse. For the
contrail to stay collimated, the flyby time must be comparable to or shorter than
the crossing time. Thus, we expect
\begin{equation}
    t_{\rm flyby} \lesssim t_{\rm cross} \approx 20\,{\rm Myr}\;,
\end{equation}
and  a speed of 
\begin{equation}
    v_{\rm flyby} \gtrsim \frac{L}{t_{\rm cross}} \approx \frac{6\,{\rm kpc}}{20\,{\rm Myr}} \approx 300\,{\rm km\,s^{-1}}\;.
\end{equation}

From Fig. \ref{fig2}, the velocity gradient of the contrail is around 10
km\,s$^{-1}$ kpc$^{-1}$, which implies a timescale of

\begin{equation}
    t_{\rm shear} \approx \frac{1}{10\,{\rm km\,s^{-1}\,kpc^{-1}}} \approx 100\,{\rm Myr}\;.
\end{equation}
Since $t_{\rm flyby} \approx t_{\rm cross} < t_{\rm shear}$, gravity from
the dark matter in the galaxy is not strong enough to bend the trajectory
significantly, thus the collimated structure is preserved.

\paragraph{Constraints on the Massive Object} 
Gravitational wake theory
\citep{2021MNRAS.503.4466L}, see also \citet{2023ApJ...945...39K,2024MNRAS.527.5503O} predicts a
mass of 
\begin{equation}
    M \sim \frac{\sigma_v v_{\rm flyby} r}{G}
\end{equation}
for the dense object, using $r =
100$~pc as the effective interaction radius and 10 km s$^{-1}$ as the velocity
dispersion of contrail, we derive
\begin{equation}
    M \sim 1\times10^7\,M_\odot \;,
\end{equation}
which matches expectations for an massive black hole or a dwarf galaxy’s central region.
If a dwarf galaxy caused the contrail, it should lie along the contrail’s projected trajectory. However, at NGC 3627’s distance (11 Mpc), a faint dwarf ($\lesssim 10^8 \,M_\odot$, V $>$ 22 mag) would require deep surveys for detection.

\paragraph{Dynamical differences and 3D structure}
The first 6 kpc of the contrail has a flat velocity gradient of about 10
km\,s$^{-1}$ kpc$^{-1}$, which is much lower than the velocity gradient of the
spiral arm (30 km\,s$^{-1}$ kpc$^{-1}$). This difference indicates that the
first part of the
contrail, which we expect to stay outside of the Galactic disk, is not
significantly affected by the galaxy's overall rotation, as compared to the
extended part, which may stay at the interface between the galactic disk and Circumgalactic medium (CGM).





\paragraph{Massive Dark Objects as the origin of Galactic Contrails} 
Such contrails serve as unique tracers of dark compact objects. The required
mass and velocity parameters overlap with predicted black hole populations
\citep{2024MNRAS.527.5503O} and dwarf galaxies  \citep{2025arXiv250621841P}. 
The contrail may also other origins: Recent JWST observations have unveiled the little red dots \citep{Geraldi2024,Marconcini2025,Akins2024,Labbe2025}, which
are considered as compact, red-tinted, early-universe galaxies, likely powered
by supermassive black holes or dense star-forming regions, and these and their counterparts can also
be the cause of this contrail.
Future deep surveys (e.g., Rubin~LSST  \citep{2020ApJ...900..139F}) could search
for stellar counterparts along the contrail axis, while higher-resolution ALMA
observations may reveal kinematic signatures of the perturbing object. \\


\section*{Acknowledgments}

GXL acknowledges support from NSFC grant Nos. 12273032 and 12033005.
We thank the teams behind the following legacy surveys for making their superb data publicly available: PHANGS-ALMA\footnote{\url{ https://phangs.stsci.edu}}, PHANGS-JWST, and THINGS\footnote{\url{ https://www.mpia.de/THINGS}}. This research relied on these foundational datasets.

\bibliographystyle{apsrev4-1}

\bibliography{oja_template}

\begin{appendix}

\end{appendix}

\end{document}